\documentclass[prd,aps,twocolumn,amsmath,amssymb,nofootinbib,preprintnumbers]
{revtex4}

\usepackage{graphicx}
\usepackage{dcolumn}
\usepackage{bm}
\usepackage{amsmath}
\usepackage{amsfonts}
\usepackage{euscript,bbm}
\usepackage{epsfig}
\usepackage{ifthen}
\usepackage{psfrag}
\usepackage{slashed}

\newcommand{\Lu}  {{\cal L}}  
\newcommand{\Br}  {{\cal B}}   

\newcommand{\tev}  {{\rm TeV}}
\newcommand{\gev}  {{\rm GeV}}
\newcommand{\invfb}{{\rm fb}^{-1}}

\newcommand{\mzero} {m_{0}}
\newcommand{\mhalf} {m_{1/2}}
\newcommand{\azero} {A_{0}}
\newcommand{\tanb} {\tan\beta}

\newcommand{ \gluino}   {\tilde{g}}
\newcommand{ \squark}[1]   {\tilde{q}_{#1}}

\newcommand{ \staupm}[1]  {\tilde{\tau}_{#1}^{\pm}}
\newcommand{ \stau}[1]  {\tilde{\tau}_{#1}}
\newcommand{ \smu}[1]  {\tilde{\mu}_{#1}}
\newcommand{ \slepton}[1]  {\tilde{\ell}_{#1}}
\newcommand{ \schizero }[1] {\tilde{\chi}_{#1}^{0}}
\newcommand{ \schipm }[1] {\tilde{\chi}_{#1}^{\pm}}

\newcommand{\met} {{E\!\!\!\!/_{\rm T}}}

\newcommand{\ptof}[1] {p_{{\rm T}, #1}}
\newcommand{\ptvisof}[1]  {p_{{\rm T}, #1}^{\rm vis}}

\newcommand{\pt} {p_{\rm T}}

\newcommand{\mtautau} {m_{\tau\tau}}
\newcommand{\mtautauEnd}{m_{\tau\tau}^{\rm end}}

\newcommand{\mtaumu}{m_{\tau\mu}}

\newcommand{\ptslope}{{\rm slope}(\ptvisof{\tau})}
\newcommand{\ptslopeHigh}{{\rm slope}(\ptvisof{\tau ({\rm high}) })}
\newcommand{\ptslopeLow}{{\rm slope}(\ptvisof{\tau ({\rm low}) })}
\newcommand{\ptslopePlus}{{\rm slope}(p_{\rm T\ +}^{\rm vis}) }
\newcommand{\ptslopeMean} {\left<\ptslope\right>}

\newcommand{ \spheno } {{\tt SPheno}}
\newcommand{ \pythia } {{\tt PYTHIA}}
\newcommand{ \pgs }    {{\tt PGS4}}

\newcommand{\deltaLFV} {\delta_{RR, {\rm LFV}} }
\newcommand{\tauh} {\tau_{\rm h} }
\newcommand{\emm} {\mathcal{M} }

\begin{document}

%
\title{Lepton Flavor Violation at the Large Hadron Collider}

\author{Rouzbeh Allahverdi$^{1}$}
\author{Bhaskar Dutta$^{2}$}
\author{Teruki Kamon$^{2,3}$}
\author{Abram Krislock$^{2,4}$}

\affiliation{$^{1}$~Department of Physics and Astronomy, University of New Mexico, Albuquerque, New Mexico 87131, USA \\
$^{2}$~Department of Physics \& Astronomy, Mitchell Institute for Fundamental Physics, Texas A\&M University, College Station, Texas 77843-4242, USA \\
$^{3}$~Department of Physics, Kyungpook National University, Daegu 702-701, South Korea \\
$^{4}$~Department of Physics, AlbaNova, Stockholm University, SE-106 91, Stockholm, Sweden}

\begin{abstract} 

We investigate a potential of discovering lepton flavor violation (LFV) at the Large Hadron Collider. A sizeable LFV in low energy supersymmetry can be induced by massive right-handed neutrinos, which can explain neutrino oscillations via the seesaw mechanism. We investigate a scenario where the distribution of an invariant mass of two hadronically decaying taus ($\tauh\tauh$) from $\schizero{2}$ decays is the same in events with or without LFV. We first develop a transfer function using this ditau mass distribution to model the shape of the non-LFV $\tauh\mu$ invariant mass. We then show the feasibility of extracting the LFV $\tauh\mu$ signal. The proposed technique can also be applied for a LFV $\tauh e$ search.

\end{abstract}
MIFPA-12-06

\maketitle



\section{Introduction}\label{intro}

Supersymmetry (SUSY)~\cite{susy} is one of the most promising candidates of physics beyond the standard model (SM). In SUSY, the gauge couplings can unify at a high scale, which leads to a successful realization of grand unified theories. Also, it can solve the gauge hierarchy problem and yield a dark matter candidate to explain the 23\% of the energy density of the Universe. However, the flavor sector of the minimal supersymmetric standard model can cause trouble due to the fact that SUSY breaking terms can induce large flavor changing neutral currents. The experimental constraints on flavor changing neutral currents require flavor degeneracy of the SUSY particles, especially for the first and second generations, if SUSY particles are lighter than around 2-3 TeV~\cite{lfvFCNCconstraints}.

In order to solve the flavor problem, the simplest assumption is that the squark and slepton masses are unified, flavor diagonal, and degenerate in the mass basis of quarks and leptons. The minimal version of this picture is the minimal supergravity framework (mSUGRA)~\cite{msugra}, which is described by 4 parameters and a sign: The universal scalar mass, $\mzero$, the universal gaugino mass, $\mhalf$, the universal triliner coupling, $\azero$, the ratio of the vacuum expectation values of the two Higgs fields, $\tanb$, and the sign of the Higgs mixing term, $\mu$, in the superpotential. These parameters are specified at the grand unified scale $M_G$. In this model, however, the renormalization group equation splits the third generation from the other two because of the large top Yukawa coupling. The Cabbibo-Kobayashi-Masukawa matrix then allows this model to have observable flavor violation involving the third generation, e.g., the branching ratio of $b\rightarrow s\gamma$.

In the leptonic sector, the neutrino oscillation data suggest that we also have a neutrino mixing matrix, the Maki-Nakagawa-Sakata-Pontecorvo matrix~\cite{mnspMatrix}. The light neutrino masses are usually generated via the seesaw mechanism, involving heavy Majorana masses for right-handed neutrinos, which are introduced as additions to the SM quarks and leptons. The precise seesaw formula for the light neutrino mass matrix with three generations is given by~\cite{seesaw}
\begin{equation}\label{eq:neutrinoMassMatrix}
\emm_{\nu} = \emm^T_D (\emm_R)^{-1} \emm_D,
\end{equation}
where $\emm_D$ is the Dirac neutrino mass matrix and $\emm_R$ is the Majorana matrix, which consists of three right-handed neutrinos ($\nu^c$) that have masses at the scale $v_{B-L}$ corresponding to a new (local) $B-L$ symmetry. The neutrino mixing angles in such schemes would arise as a joint effect from two sources: (i) mixings among the right-handed neutrinos present in $\emm_R$ and (ii) mixings among different generations present in the Dirac mass matrix $\emm_D$. The (physical) neutrino oscillation angles will also receive contributions from mixings among the charged leptons.

The neutrino flavor mixings induced by the seesaw mechanism (as needed by oscillation data) can generate lepton flavor violation (LFV) effects. Within the SM, extended minimally to accommodate the seesaw mechanism, such effects are extremely small in any process due to a power suppression factor $(1/v_{B-L})^2$, as required by the decoupling theorem. However, this situation is quite different if there is low energy SUSY. The suppression factor for LFV effects then becomes much weaker, taking the value $(1/M_{\rm SUSY} )^2$. This can lead to observable LFV effects at low energies, as noted in a number of papers~\cite{lfvObservableEffects}. The main difference is that, with low energy SUSY, LFV can be induced in the slepton sector, which can then be transferred (through one loop diagrams involving the exchange of gauginos) to the leptons suppressed only by a factor $(1/M_{\rm SUSY} )^2$. The experimental evidence for neutrino oscillation is thus a strong indicator that there might very well be LFV, assuming the validity of low energy SUSY. Searches for LFV processes such as $\tau\rightarrow\mu\gamma$ and/or $\mu\rightarrow e\gamma$ can therefore be an important source of information on the $\nu^c$ mixings in $\emm_R$ and/or family mixings in $\emm_D$. 

Motivated by this natural occurrence of LFV due to neutrino oscillation, in this paper, we study the prospect of discovering LFV in the slepton sector at the Large Hadron Collider (LHC) using mass reconstruction techniques. We do not use any particular scenario to generate the LFV in the leptonic sector. Instead, we use a bottom up approach so that this study can be applied to any model of LFV. 

At the LHC, the squarks and gluinos should be produced from the proton-proton collisions in any SUSY model if the masses of these colored objects are $\lesssim 3~\tev$. These colored objects decay via a cascade process ultimately into the lightest SUSY particle, which is the lightest neutralino ($\schizero{1}$) in this paper. We consider the well motivated scenario where R-parity is conserved, so that the $\schizero{1}$ is stable and can be the dark matter candidate. At the detector, the $\schizero{1}$ escapes detection and forms missing energy. In order to be able to see the LFV from the cascade decays, the cascade decay chain must contain sleptons. This happens in a large class of SUSY models where the squarks and gluinos decay into $\schizero{2}$ or the lightest chargino, $\schipm{1}$, by emitting quarks (jets). The $\schizero{2}$ and the $\schipm{1}$ then decay into sleptons and leptons. The sleptons eventually decay into leptons and the $\schizero{1}$. The entire decay process is:
\begin{equation}\label{eq:decayProcess}
\gluino \rightarrow q\squark{} \rightarrow q q \schizero{2} \rightarrow q q \ell \slepton{}\rightarrow q q \ell \ell \schizero{1}.
\end{equation} 
It is possible that $\schipm{1}$ may be produced instead of $\schizero{2}$, in which case $\schipm{1}$ decays into $\ell \nu \schizero{1}$ via charged slepton or sneutrino production. 

In the mSUGRA framework, the first two generation sleptons, $\tilde{e}$ and $\tilde{\mu}$, are highly degenerate without LFV effects. However, introducing some LFV effects causes a mass splitting between these SUSY particles. If these particles are accessible at the LHC, then measuring this mass splitting can help to probe the LFV~\cite{lfvWithSleptonMassSplitting}. However, the first two generation sleptons are very difficult to probe by direct production at the LHC. Additionally, for larger $\tanb$, the staus are much lighter compared to the first two generation sleptons. Thus, the branching ratio into these heavier sleptons from other SUSY particles is very small. We propose a new technique in this paper that only depends on a mass measurement of the lightest slepton.

Our final state of interest contains at least two hadronically decaying taus, $\tauh$, plus missing transverse energy, $\met$. Using the $2 j + 2\tauh + \met$ signal, it has been shown in various studies~\cite{HinchliffePaige, darkMatterWithDitau, sscOverabundanceRegion, nuSUGRASignals, MirageSignals} that the $\tauh\tauh$ invariant mass distribution, $\mtautau$, involving opposite sign (OS)$-$like sign (LS) combinations of the two taus shows a clear endpoint or peak, which is a function of the lighter stau mass and the two lightest neutralino masses. In particular, it is also shown that using kinematic observables (such as various invariant mass distributions of the jets and taus as well as the $p_T$ distribution of the taus), one can solve for the gluino, squark, lighter stau, and lightest neutralino masses in various SUSY models~\cite{darkMatterWithDitau}. Since the masses are reconstructed from the observables, this technique is very general and can be applied to any generic SUSY model.

Since the slepton-neutralino and slepton-chargino interactions carry the information of LFV, we investigate one of the major cascade decay chains involving the sleptons. As we have discussed above, in non-LFV scenarios, the final states contain only taus when staus are the lighter of sleptons, and the $\mtautau$ distribution shows a clear endpoint and peak. However, when LFV is present, we can have the decay modes $\schizero{2} \rightarrow \tau \mu \schizero{1}$ and/or $\schizero{2} \rightarrow \tau e \schizero{1}$. The authors of \cite{complementaryLFV} proposed a search for the LFV signal in an excess in OS $\mu$-$\tauh$ over OS $e$-$\tauh$ events. Their analysis assumed no LFV decays in the $e$-$\tauh$ channels. In this paper, we propose a complementary method using a ``transfer'' function, which can be used for the $\mu$-$\tauh$ and $e$-$\tauh$ LFV channels simultaneously. (see Sec.~\ref{lfvSearch}).

The paper is organized as follows.
In Sec.~\ref{origin} we discuss further the origins of LFV terms in SUSY, as well as our particular implementation of such LFV terms for this study.
In Sec.~\ref{massDetermination} we show ``measurements'' of the SUSY masses using non-LFV decays. Measuring these SUSY masses is a crucial step in estimating the non-LFV background.
In Sec.~\ref{lfvSearch} we propose a technique to estimate this background and extract the effects of LFV. This technique is based upon the hadronic tau pair signal that is common to both the LFV and non-LFV case. We report the feasibility of detecting the LFV component at the LHC. 
Finally, we present our conclusions in Sec.~\ref{conclu}.

\section{Origin and Implementation of LFV}
\label{origin}

LFV effects can be generated by the neutrino seesaw mechanism in SUSY as follows. At the grand unified scale we have the mSUGRA boundary condition and thus, there is no flavor violation anywhere except in the Yukawa couplings. It should be noted that in the absence of neutrino masses, there is only one leptonic Yukawa matrix, $Y_l$, for the charged leptons, which can be diagonalized at $M_G$ and which will remain diagonal to the weak scale. Thus, it would not induce any flavor violation in the slepton sector in mSUGRA models. However, to satisfy the neutrino mixing data, the right-handed neutrinos have masses of order $v_{B-L}$, which is lower than $M_G$. These right-handed neutrino masses must be $\sim10^{12}-10^{15}~\gev$. In the momentum regime $v_{B-L}\leq \mu \leq M_G$ where the $\nu^c$ fields are active, the soft masses of the sleptons will feel the effects of LFV in the neutrino Yukawa sector through the renormalization group evolution. At the scale $v_{B-L}$, the slepton mass matrix is no longer universal in flavor, and this nonuniversality will remain down to the weak scale. 

Since we are interested in the phenomenological aspects of such LFV at the LHC, we simply introduce a term which causes LFV by hand within the charged slepton mass matrix. The charged slepton mass matrix is a $6\times6$ matrix and is given by 
\begin{equation}\label{eq:sleptonMassMatrix}
 \emm^2_{\slepton{}} = \left(
	\begin{array}{cc}
	\emm^2_{LL} & \emm^2_{LR} \\
	\emm^2_{LR} & \emm^2_{RR}
	\end{array}
  \right),
\end{equation} 
where $\emm_{LL}^2$ represents the $3\times3$ matrix for the soft masses for left sleptons, $\emm_{RR}^2$ represents the $3\times3$ matrix for the soft masses for right sleptons, and $\emm_{LR}^2$ represents the $3\times3$ diagonal matrix with elements $m_l(A_l+\mu\tan\beta)$, where $A_l$ is the trilinear soft mass term and $m_l$ is the diagonal charged lepton mass ($Y_l v_d$) for generation $l$. In mSUGRA, 
\begin{equation}\label{eq:msugraUnifiedMass}
  \emm_{LL}^2=\emm_{RR}^2=m_0^2\left(
  \begin{array}{ccc}
	1 & 0 & 0 \\
	0 & 1 & 0 \\
	0 & 0 & 1
  \end{array}
  \right)
\end{equation} and $A_l=A_0$.
The effects of LFV can be produced when off-diagonal elements in $\emm_{LL,LR,RR}^2$ are present. 
During the diagonalization of this mass matrix (which puts us in the mass eigenstate basis of sleptons), if there is no such off-diagonal LFV element in $\emm_{LL,LR,RR}^2$, then the mass eigenstates are states of pure flavors. However, the off-diagonal LFV element causes the mass eigenstates to become mixtures of different flavors. These mixed-flavor mass eigenstates naturally act sometimes as one flavor, and sometimes as another.

Consequently, if the second lightest neutralino, $\schizero{2}$, is produced at the LHC from the cascade decays of squarks, then it can decay to $\tau\tau$, $\mu\mu$, or $ee$ final states, plus $\met$ due to the undetectable lightest neutralino, $\schizero{1}$. On the other hand, when LFV producing off-diagonal elements are introduced, the final states from the $\schizero{2}$ decay can also have $\tau\mu$, $\mu e$, or $e\tau$ final states plus $\met$. If, for example, the $(2, 3)$ element [which is the same as the $(3, 2)$ element] of any or all of the $\emm^2_{LL,LR,RR}$ are nonzero then $\tau\mu$ final states plus $\met$ will appear in the LFV $\schizero{2}$ decay. In this paper, we study LFV by introducing a nonzero value of the $(2, 3)$ element of the $\emm_{RR}^2$ matrix. This new element also will allow the $\tau\rightarrow\mu\gamma$ decay and therefore is constrained, i.e., $\Br(\tau\rightarrow\mu\gamma)\leq 4.4 \times 10^{-8}$~\cite{BRtaumugamma}. We will define 
\begin{equation}\label{eq:deltaLFV}
  \deltaLFV = \frac{\left[\emm^2_{RR}\right]_{23}}{\left[\emm^2_{RR}\right]_{33}}
\end{equation} 
to quantify the amount of LFV. This quantity will enter into the LFV decay modes of neutralinos and sleptons at the LHC and $\tau\rightarrow\mu\gamma$ amplitude.

\section{Determining the Masses of $\stau{1}$, $\schizero{1}$, and $\schizero{2}$} \label{massDetermination}

As mentioned in the Introduction, we see that the LFV and non-LFV decay channels both involve the $\stau{1}$, $\schizero{1}$, and $\schizero{2}$. In fact, for LFV which is not too large to introduce appreciable change in the stau mass, it is possible to reconstruct the $\tauh\mu$ invariant mass distribution, $\mtaumu$, whose endpoint will coincide with the end point of the $\mtautau$ distribution. However, the $\mtaumu$ distribution can also be present even when there is non-LFV, due to the leptonic tau decay, $\tau\rightarrow\nu\bar\nu\mu$. This is a major background to the LFV signal. 

Thus, in order to understand the LFV signal at the LHC, we must first be able to estimate this background. In order to do this, we must measure the masses of the SUSY particles involved in both the LFV and non-LFV signals as accurately as possible. Thus, in this section we demonstrate the technique used to determine the SUSY particle masses involved in the LFV decay chain, which we will discuss in Sec.~\ref{lfvSearch}. Specifically, we will determine the $\stau{1}$, $\schizero{1}$, and $\schizero{2}$ masses.

Since the subsystem $\schizero{2}$-$\stau{1}$-$\schizero{1}$ involves the decay chain that is essential for the study of LFV at the LHC, we need to produce this subsystem in the cascade decays of $\squark{}$, $\gluino$ that can occur as follows:
\begin{equation}
	\squark{L} \rightarrow q  \schizero{2} \rightarrow q  \tau^{\mp}  \staupm{1}
		\rightarrow q  \tau^{\mp}  \tau^{\pm}  \schizero{1}
\label{eq:regularDecay}
\end{equation}
The signal of this decay chain at the LHC is characterized by high energy jets (from the squark decays), a pair of oppositely charged tau leptons, and a large missing energy signal (from the lightest neutralino, which escapes detection). We need to determine the $\schizero{2,1}$ and $\stau{1}$ masses from this chain.

In order to determine the masses, we  choose a  model in mSUGRA~\cite{msugra}, generate the mass spectrum for that model using \spheno~\cite{spheno}, simulate LHC collision events at $\sqrt{s} = 14~\tev$ using \pythia~\cite{pythia}, and model the detector response with \pgs~\cite{pgs}. Although the model point is excluded by the LHC experiments, we choose $\mzero = 250~\gev$, $\mhalf = 350~\gev$, $\azero = 0$, $\tanb = 40$, and $\mu > 0$ as a benchmark point for a comparison with our previous study for the coannihilation case ~\cite{darkMatterWithDitau}. The relevant masses $\schizero{2,1}$ and $\stau{1}$ at this benchmark point are shown in Table~\ref{tab:massBenchmark}. We stress that the technique presented here can be used for any SUSY model point with heavier SUSY masses as long as the $\schizero{2}$-$\stau{1}$-$\schizero{1}$ subsystem is present in the cascade decay chains of SUSY particles.

As stated above, our benchmark point has already been ruled out at the LHC, since $\gluino \sim \squark{L} \simeq 800~\gev$~\cite{LHCexclusionATLASandCMS}. However, since the $\squark{}$ and $\gluino$ masses set the cross section of SUSY production in the analysis, we can go to any other point of the parameter space with larger $\squark{}$ and $\gluino$ masses. These larger masses do not change the presence of the $\schizero{2}$-$\stau{1}$-$\schizero{1}$ subsystem of interest. Thus, by scaling the luminosity by the cross section times the branching ratio of $\squark{}$ and $\gluino$ into $\schizero{2}$, we can achieve the same degree of accuracies for determining the $\schizero{2,1}$ and $\stau{1}$ masses or the amount of LFV. For this parameter space point $\Br(\squark{L} \rightarrow \schizero{2}q)\simeq0.35$ and $\Br(\squark{R} \rightarrow \schizero{2}q)\simeq0$. However, we can change these branching ratios by changing the wino-bino content of the lightest two neutralinos. For instance, by departing from the gaugino mass unification scheme in mSUGRA, we can find a model where the $\schizero{2}$ is almost entirely bino-like. In this case, $\Br(\squark{R} \rightarrow \schizero{2}q)\simeq1$, increasing the overall branching ratio of squarks into our decay chain of interest by a factor of about three.

Thus, the luminosity requirement in the analysis of the $\squark{L}$ chain is dictated by $\sigma (\squark{L},\gluino) \Br(\squark{L},\gluino \rightarrow \schizero{2}) A_{j+\met} A_{\tau\tau}$, where $\sigma (\squark{L},\gluino)$ is the production cross section of $\squark{L}$ and $\gluino$ at the LHC, $A_{j+\met}$ is the acceptance for the jets plus $\met$ system, and $A_{\tau\tau}$ is acceptance for the $\tau\tau$ system. Now, as the $\squark{L}$ and $\gluino$ masses increase, $\sigma (\squark{L},\gluino)$ will go down. However, $A_{j+\met}$ and $A_{\tau\tau}$ can be maintained the same (by adjusting the cuts). Thus, to achieve the same result for a different $\sigma \times \Br$, the analysis technique remains the same; only a different amount of luminosity is needed. 

Our benchmark point described above (and shown in Table~\ref{tab:massBenchmark}) has an overall production cross section of $6.6$ pb according to our \pythia\ simulation. If we choose a model point that has a similar decay chain scenario and that has not already been ruled out, we can estimate how much more luminosity would be required as compared to our benchmark. For example, another mSUGRA point (with $\mzero = 410~\gev$, $\mhalf = 750~\gev$, $\azero = 0$, $\tanb = 40$, and $\mu > 0$) has larger squark and gluino masses, yet still has the $\tau\tau$ system of interest. This point has a cross section of $0.094$ pb. Thus, we would naively expect to require a factor of 70 more luminosity than we present here. As suggested above, the acceptances may also change. However, we can alter the cuts to maintain the luminosity requirement. We have shown a similar luminosity scaling behavior in a previous analysis~\cite{MirageSignals}

\begin{table}
\caption{Relevant mass spectrum for our chosen mSUGRA benchmark point: $\mzero = 250~\gev$, $\mhalf = 350~\gev$, $\azero = 0$, $\tanb = 40$, and $\mu > 0$. All masses are in GeV.}
\label{tab:massBenchmark}
\begin{center}
\begin{tabular}{cc} \hline \hline
   Particle & True mass  \\ \hline
   $\stau{1}$ & $186.7$  \\
   $\schizero{1}$ & $141.5$  \\
   $\schizero{2}$ & $265.8$
 \\ \hline \hline
\end{tabular}
\end{center}
\end{table}


Our analysis proceeds as follows. In order to select our SUSY events from the background of other SUSY events and SM background events, we employ similar cuts as were used in \cite{darkMatterWithDitau}. For us to select the event for analysis, it must have the following:
\begin{itemize}
	\item at least two hadronically decaying tau leptons with $\ptvisof{\tau} \ge 15~\gev$~\cite{CMStau},
	\item at least two jets, where the leading two jets have $\ptof{{\rm jet}1,2} \ge 100~\gev$,
	\item missing transverse energy, $\met \ge 200~\gev$, and
	\item scalar sum, $h_{\rm T} = \met+\ptof{{\rm jet}1}+\ptof{{\rm jet}2} \ge 600~\gev$.
\end{itemize}

Once events have been selected in this way, we select all pairs of tau leptons from each event. Each pair is characterized as either LS or OS based upon the reconstructed charge of the taus in the pair. To remove the combinatoric background of incorrect combinations of taus, we can perform the OS$-$LS subtraction for any kinematical distribution in which we are interested. Doing this, we construct the following kinematical distributions:
\begin{itemize}
	\item the $\tauh\tauh$ invariant mass distribution, $\mtautau$,
	\item the distribution of the transverse momentum of the higher $\pt$ tau, $\ptvisof{\tau ({\rm high}) }$,
	\item the distribution of the transverse momentum of the lower $\pt$ tau, $\ptvisof{\tau ({\rm low}) }$, and
	\item the distribution of the transverse momentum sum of the two taus, $\ptslopePlus = \ptvisof{\tau ({\rm high}) }+ \ptvisof{\tau ({\rm low}) }$.
\end{itemize}

In order to determine the $\stau{1}$, $\schizero{1}$, and $\schizero{2}$ masses, we need three independent observables. Here, we over constrain the system with four observables, which allows us to reduce the uncertainty in the measurement. The four observables we choose are as follows.

\begin{figure}
	\centering
	\includegraphics[width = 8cm]{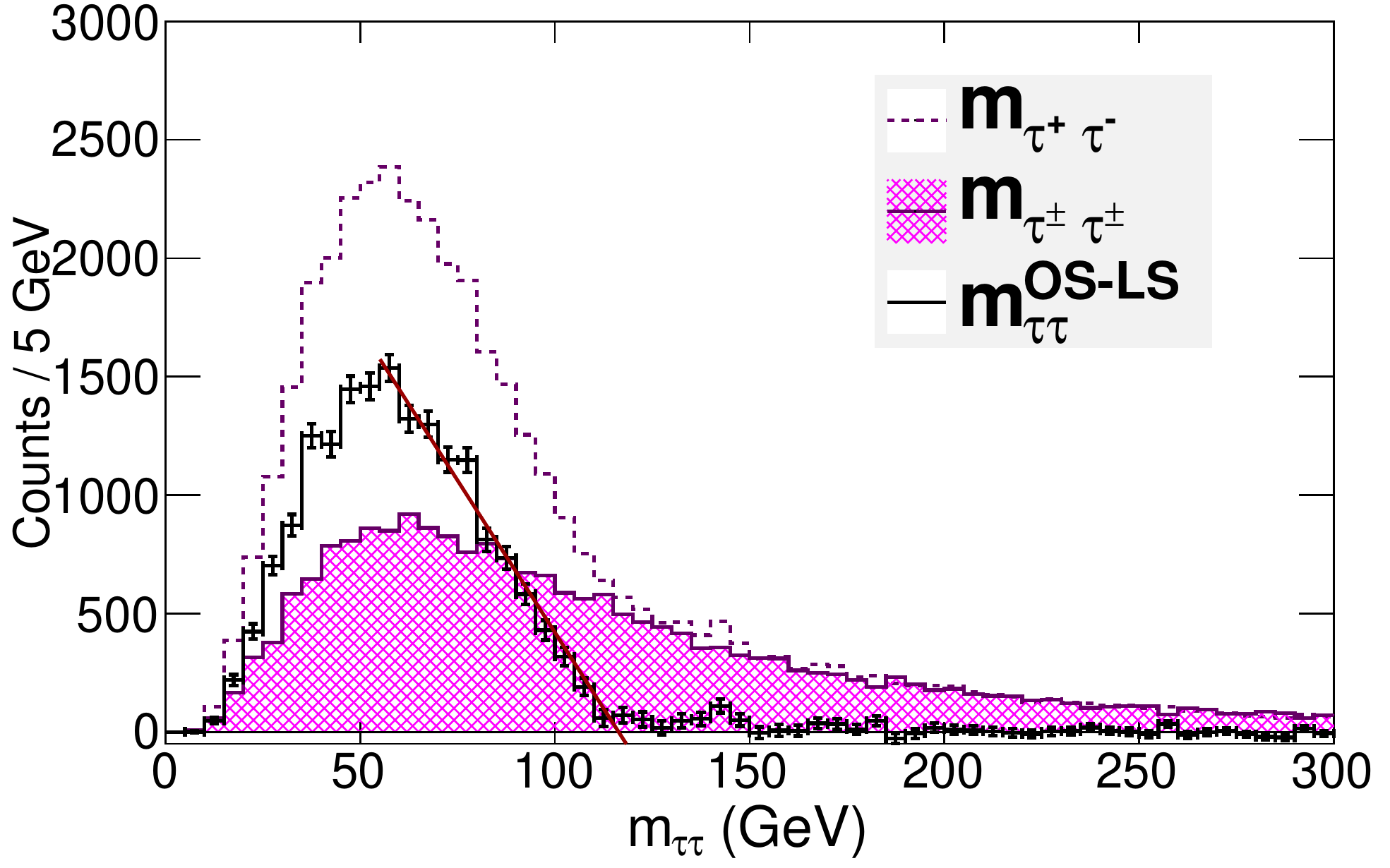}
	\caption{The $\tauh\tauh$ invariant mass distribution for our benchmark point, shown in Table~\ref{tab:massBenchmark}. A linear fit finds the endpoint of the distribution. This distribution represents an integrated luminosity of $1000~\invfb$. However, we also report the situation for a lower luminosity of $300~\invfb$ in this paper.}
	\label{fig:ditauBenchmarkEndpoint}
\end{figure}

The $\tauh\tauh$ invariant mass distribution, $\mtautau$, has a maximum value for taus coming from the decay chain shown in Eq.~(\ref{eq:regularDecay}). (We select taus from this decay chain using the OS$-$LS technique described above). This maximum value, $\mtautauEnd$, depends on all three masses: $\mtautauEnd = f_{1} (m_{\stau{1}}, m_{\schizero{1}}, m_{\schizero{2}})$. A sample $\tauh\tauh$ invariant mass distribution is shown in Fig.~\ref{fig:ditauBenchmarkEndpoint}.

The slope of the log scale plotted $\pt$ distributions can also be made into observables. For instance, the slope of the higher $\pt$ tau is a function of all three masses: $\ptslopeHigh = f_{2} (m_{\stau{1}}, m_{\schizero{1}}, m_{\schizero{2}})$. Also, the average of the slopes of the high and low taus, $\ptslopeMean = \frac{1}{2} \left( \ptslopeLow + \ptslopeHigh \right)$, is a function of all three masses as well: $\ptslopeMean = f_{3} (m_{\stau{1}}, m_{\schizero{1}}, m_{\schizero{2}})$. Lastly, the slope of the transverse momentum sum distribution is a function of two of the masses: $\ptslopePlus = f_{4} (m_{\schizero{1}}, m_{\schizero{2}})$.

In order to find these functional forms, $f_{1}$, $f_{2}$, $f_{3}$, and $f_{4}$, we vary the masses near our benchmark point. We change each of the masses, $m_{\stau{1}}$, $m_{\schizero{1}}$, and $m_{\schizero{2}}$, in turn while holding the others constant. By repeating the simulation for each of these varied points, we can find the functional forms of each observable as a function of one mass. The overall functional forms are then estimated by combining these one-dimensional functions into a three-dimensional function in a multiplicative way. Similar mass determination techniques (using kinematical observables and functional forms) have been demonstrated before. See for example \cite{HinchliffePaige, darkMatterWithDitau, sscOverabundanceRegion, nuSUGRASignals, MirageSignals}.

Once the functional forms are found in this way, we can invert them to solve for all three masses. In principle, this can be done algebraically. However, such a system of equations is complicated and over constrained. Instead, we invert these equations by use of the Nelder Mead method. This method is a nonlinear optimization technique that we employ to search for the masses that best fit the observables according to the functional forms. This is nearly identical to the method used in \cite{MirageSignals}. We find two solutions for the SUSY masses using this method, due to the nonlinear nature of the functional forms. The results of our mass determination are shown in Table~\ref{tab:massMeasurements}.

\begin{table}
\caption{Mass measurements for our chosen mSUGRA benchmark point: $\mzero = 250~\gev$, $\mhalf = 350~\gev$, $\azero = 0$, $\tanb = 40$, and $\mu > 0$. The statistical uncertainties shown first are estimated for a luminosity of $\Lu = 1000(300)~\invfb$. The systematic uncertainty due to a jet energy scale mismeasurement of 3\%~\cite{systematic} has also been estimated, shown second. All masses are in GeV.}
\label{tab:massMeasurements}
\begin{center}
\begin{tabular}{ccc} \hline \hline
   Particle mass & Solution one & \hspace{0.1cm}Solution two \\ \hline
   $\stau{1}$ : $186.7$ & 
   	$181.5 \pm 3.7(5.1) \pm 4.1$ & 
	$205.8 \pm 5.9(6.1) \pm 5.7$ \\
   $\schizero{1}$ : $141.5$ & 
   	$140.6 \pm 5.4(6.5) \pm 6.2$ & 
	$151.4 \pm 6.4(8.6) \pm 6.3$ \\
   $\schizero{2}$ : $265.8$ & 
   	$265.3 \pm 6.2(8.5) \pm 7.3$ & 
	$278.9 \pm 9.2(11.7) \pm 9.0$
 \\ \hline \hline
\end{tabular}
\end{center}
\end{table}

\section{Searching for the LFV Signal}\label{lfvSearch}

Now that the masses have been determined (in spite of having two mass solutions), we can investigate the effects of including the $\deltaLFV$ term into our model. To see the effects of this at the LHC, we choose a value for $\deltaLFV$ and use it to rediagonalize the slepton mass matrix for our benchmark point. The lightest slepton, $\slepton{1}$, then becomes a linear combination of $\smu{R}$ in addition to the original $\stau{R}$ and $\stau{L}$ states. This allows for the LFV decays
\begin{equation}\label{eq:neutralinoTwoLFVDecay}
  \schizero{2} \rightarrow \mu \slepton{1}
\end{equation} and \begin{equation}\label{eq:sleptonLFVDecay}
  \slepton{1} \rightarrow \mu \schizero{1}.
\end{equation}
Thus, the final state may include one or more muons from the LFV decays instead of the taus in the decay chain shown in Eq.~(\ref{eq:regularDecay}). Thus, if we can select these LFV muons, we may be able to see the effect of $\deltaLFV$.

However, this analysis is complicated greatly by the fact that some of the taus decay naturally to muons. In order to see the effects of $\deltaLFV$, we need to discriminate between the muons from $\tau$ decays and the muons from the LFV decays shown in Eqs.~(\ref{eq:neutralinoTwoLFVDecay}) and (\ref{eq:sleptonLFVDecay}).

Once we have found such a method of discrimination, we should be able to see the LFV signal in a kinematic observable similar to $\mtautau$. We plan to see the signal using the $\tauh\mu$ invariant mass distribution, $\mtaumu$. We start by generating the models for a few values of $\deltaLFV$. The consequences for various values of $\deltaLFV$ are shown in Table~\ref{tab:deltaLFVChanges}. In this table, we show how the lightest slepton mass changes (which we calculate by rediagonalizing the slepton mass matrix), as well as the change in the branching ratio for the decay shown in Eq.~(\ref{eq:sleptonLFVDecay}). (We calculate the tree-level decay width to determine the branching ratios.) In practice, we change the model only by introducing the additional decay channels shown in Eqs.~(\ref{eq:neutralinoTwoLFVDecay}) and (\ref{eq:sleptonLFVDecay}) (even though the former has a negligible branching ratio). However, we do not bother to change the stau mass due to the effect of $\deltaLFV$, since it shifts only slightly (easily within $1\sigma$ of our measured value in Table~\ref{tab:massMeasurements}).
\begin{table}
\caption{The effects of $\deltaLFV$ upon our benchmark point (shown in Table~\ref{tab:massBenchmark}). The LFV decay $\schizero{2} \rightarrow \mu \slepton{1}$ has a negligible decay branching ratio ($\Br \lesssim 10^{-4}$), which we do not show here. However, this decay is still included in our simulations. We note that values of $\deltaLFV$ larger than $\sim15\%$ violate the bound on the branching ratio $\Br(\tau\rightarrow\mu\gamma)\leq 4.4 \times 10^{-8}$~\cite{BRtaumugamma} for our benchmark model.}
\label{tab:deltaLFVChanges}
\begin{center}
\begin{tabular}{ccc} \hline \hline
    $\deltaLFV$(\%) & $m_{\slepton{1}}~(\gev)$ & $\Br(\slepton{1} \rightarrow \mu \schizero{1})$ \\ \hline
    0 & $186.7$ & 0 \\
    2 & $186.3$ & $4.9\times10^{-4}$ \\
    5 & $186.0$ & $3.1\times10^{-3}$ \\
    10 & $185.1$ & $1.2\times10^{-2}$ \\
    15 & $183.5$ & $2.6\times10^{-2}$ \\
  \hline \hline
\end{tabular}
\end{center}
\end{table}

We simulate the LHC signals of the models for the four nonzero values of $\deltaLFV$ using \pythia\ and \pgs\ as before. We include the branching ratios for these LFV decays in the input files given to \pythia. These four simulations are treated as realities that we may see at the LHC. Thus, we treat them as ``LHC data'', and will refer to them as such. The mass determination techniques of Sec.~\ref{massDetermination} will result in the masses found in Table~\ref{tab:massMeasurements} for each of these four LHC data points.

As stated above, to find the LFV signal, we must estimate the background from muons that naturally arise from leptonic tau decays. To do this, we simulate a point with $\deltaLFV = 0$. However, this simulation is based upon the mass determination from Sec.~\ref{massDetermination}. Thus, instead of simulating the true benchmark point for the $\deltaLFV = 0$ case, we instead choose a point that matches the values for the measured masses shown in Table~\ref{tab:massMeasurements}. In order to see the effect of the uncertainties shown for the masses in that table, we also simulate points that are $1\sigma$ away from the central values of the measured masses. This gives us a collection of points that we refer to as ``LHC simulated'' $\deltaLFV = 0$ points.

First we analyze the LHC simulated $\deltaLFV = 0$ points to understand the shape of the $\mtaumu$ distribution for the case of non-LFV. To do this, we form both the $\mtautau$ distribution (as we did above in Sec.~\ref{massDetermination}) as well as the $\mtaumu$ distribution. To form the $\mtaumu$ distribution, we select events that satisfy similar cuts as above:
\begin{itemize}
	\item at least one hadronically decaying tau lepton with $\ptvisof{\tau} \ge 15~\gev$,
	\item at least one muon with $\ptof{\mu} \ge 20~\gev$,
	\item at least two jets, where the leading two jets have $\ptof{{\rm jet}1,2} \ge 100~\gev$,
	\item missing transverse energy, $\met \ge 200~\gev$, and
	\item scalar sum, $h_{\rm T} = \met+\ptof{{\rm jet}1}+\ptof{{\rm jet}2} \ge 600~\gev$.
\end{itemize}

In order to understand the shape of the non-LFV $\mtaumu$ distribution, we relate it to the shape of the $\mtautau$ distribution using a {\it transfer function}. 


With a given small branching ratio for the LFV decay,
we take advantage of the coexistence of non-LFV and LFV decays
to construct the transfer function in the following experimental steps:
(1) We  observe and measure the $\mtautau^{\rm{non-LFV}}$ shape in our LHC data point.
(2) Using the LHC simulated $\deltaLFV = 0$ simulation (where the $\mtautau^{\rm{non-LFV}}$
	shapes in LHC data and LHC simulated are matched)
	we perform empirical fits to the $\mtautau^{\rm{non-LFV}}$ and 
	$\mtaumu^{\rm{non-LFV}}$ distributions.
(3) The transfer function is the ratio of these fit functions.

\begin{figure}
	\centering
	\includegraphics[width = 8cm]{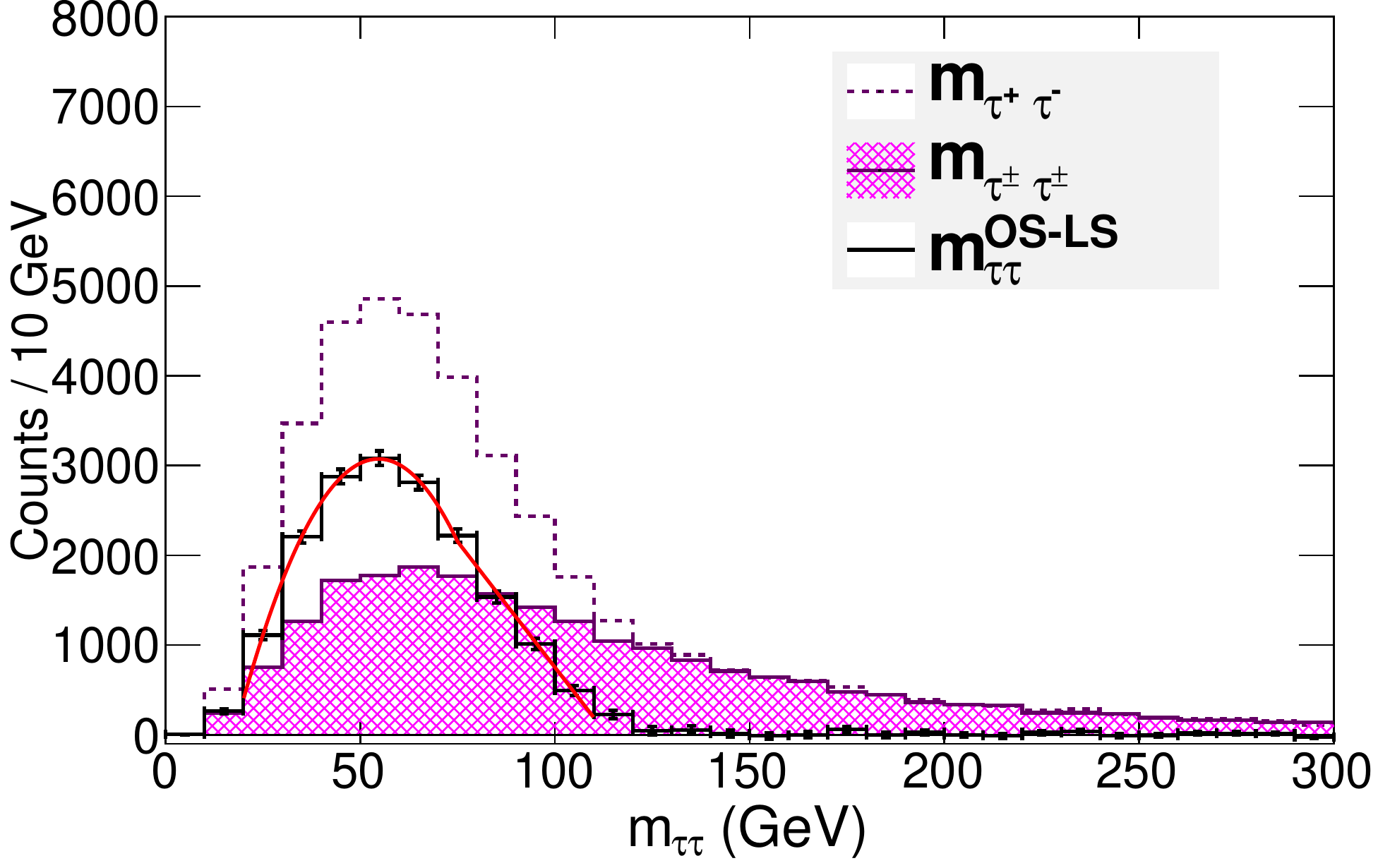}
	\caption{The $\tauh\tauh$ invariant mass distribution for our LHC simulated $\deltaLFV = 0$ point. To generate this plot, we use the first solution given in Table~\ref{tab:massMeasurements}. The shape of this non-LFV distribution is found by the fit function given in Eq.~(\ref{eq:empiricalFitFunction}). This distribution represents an integrated luminosity of $1000~\invfb$. However, we also report the situation for a lower luminosity of $300~\invfb$ in this paper.}
	\label{fig:ditauEmpiricalFit}
\end{figure}

\begin{figure}
	\centering
	\includegraphics[width = 8cm]{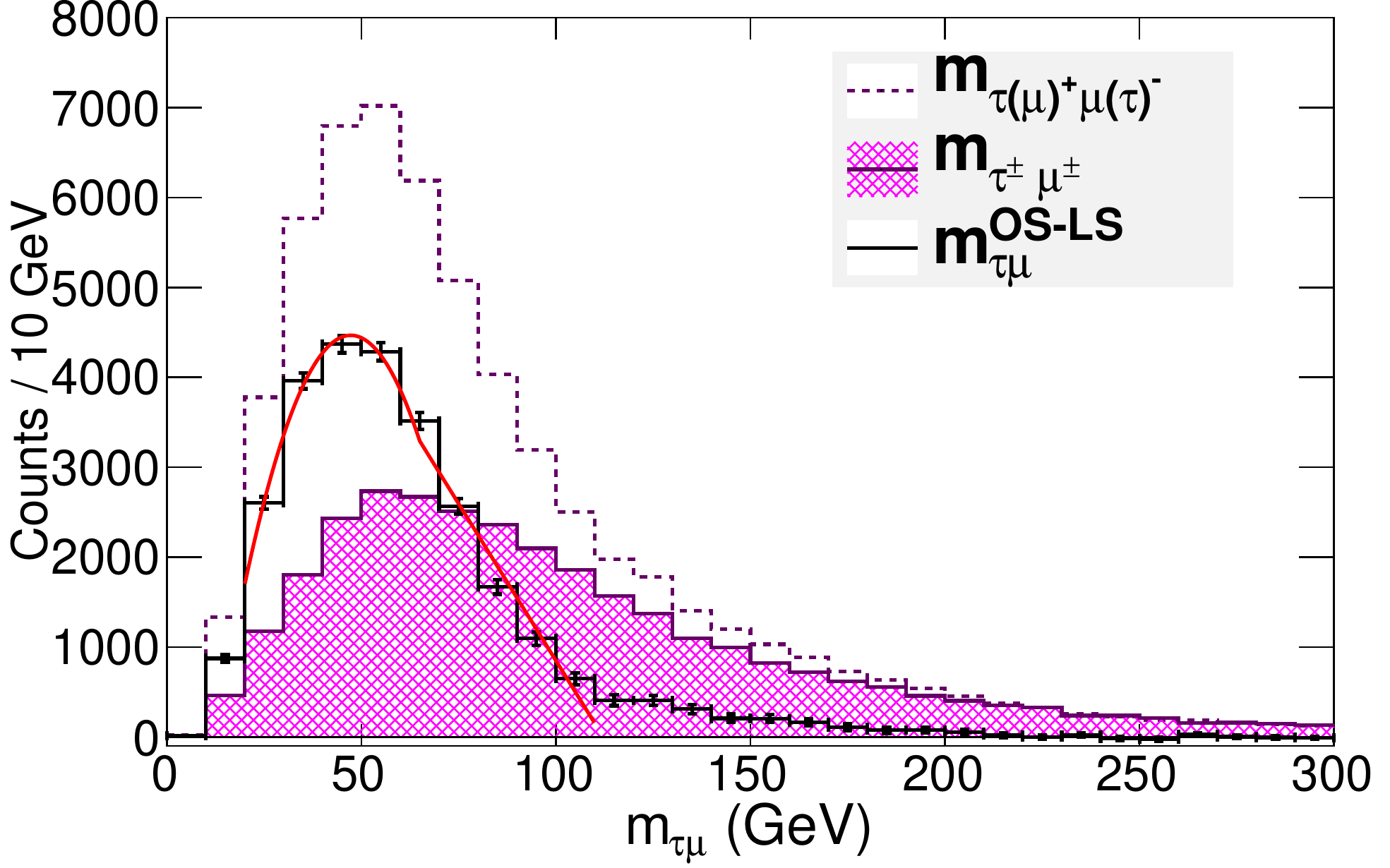}
	\caption{The $\tauh\mu$ invariant mass distribution for our LHC simulated $\deltaLFV = 0$ point. To generate this plot, we use the first solution given in Table~\ref{tab:massMeasurements}. The shape of this non-LFV distribution is found by the fit function given in Eq.~(\ref{eq:empiricalFitFunction}). This distribution represents an integrated luminosity of $1000~\invfb$. However, we also report the situation for a lower luminosity of $300~\invfb$ in this paper.}
	\label{fig:taumuEmpiricalFit}
\end{figure}

The fits that form the transfer function are shown for the $\mtautau$ and $\mtaumu$ distributions in Figs.~\ref{fig:ditauEmpiricalFit} and~\ref{fig:taumuEmpiricalFit}. These fits are used to try to minimize the statistical fluctuations by fitting with a ``smooth'' function. 
The empirical fit function we use is given by
\begin{equation}\label{eq:empiricalFitFunction}
  f(m) = \left \{
  	\begin{array}{cc}
		p_0 + p_1(m - p_3) + p_2(m - p_3)^2 & {\rm if}\ m < p_3\\
		p_0~\frac{m - p_4}{p_3 - p_4} & {\rm if}\ m \ge p_3
	\end{array},
  \right.
\end{equation}
where $m$ is the invariant mass that we are fitting, and the $p_i$s are fit parameters. We find for these distributions that the fits perform best if we fix the value of the transition parameter $p_3$. For the $\tauh\tauh$ invariant mass, we choose $p_3 = 75~\gev$, and for the $\tauh\mu$ invariant mass, we choose $p_3 = 65~\gev$. The fit range we use is $20~\gev < m < 110~\gev$. 

\begin{figure}
	\centering
	\includegraphics[width = 8cm]{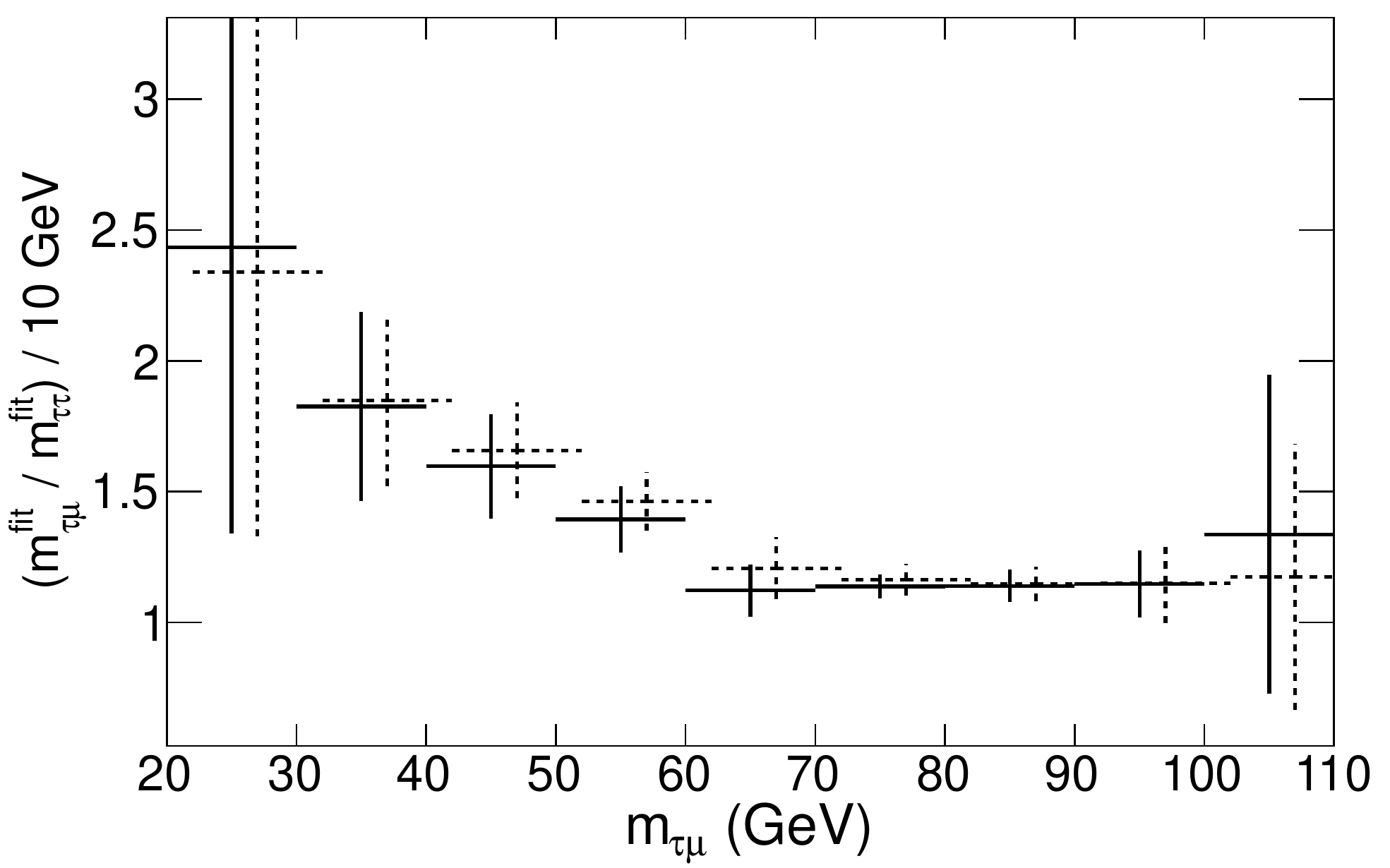}
	\caption{The transfer function for our two LHC simulated $\deltaLFV = 0$ points, each based upon one of our two SUSY mass solutions shown in Table~\ref{tab:massMeasurements}. The first solution is the solid black data points, and the second solution is the dashed black data points (which are shifted horizontally slightly for visibility). Notice that the transfer function does not vary so much between the two SUSY mass solutions. These transfer functions are generated using an integrated luminosity of $1000~\invfb$. However, we also report the situation for a lower luminosity of $300~\invfb$ in this paper.}
	\label{fig:transferFunction}
\end{figure}

The transfer function, which is formed by the ratio of these fits, is shown in Fig.~\ref{fig:transferFunction}. For the LFV points, we use the transfer function to transform the $\mtautau^{\rm{non-LFV}}$ distribution into a $\mtaumu^{\rm{non-LFV}}$ shape. Then, we subtract the $\mtaumu^{\rm{non-LFV}}$ shape from the $\mtaumu^{\rm{data}}$ distribution. Any significant excess after this subtraction makes up our LFV signal.

\begin{figure}
	\centering
	\includegraphics[width = 8cm]{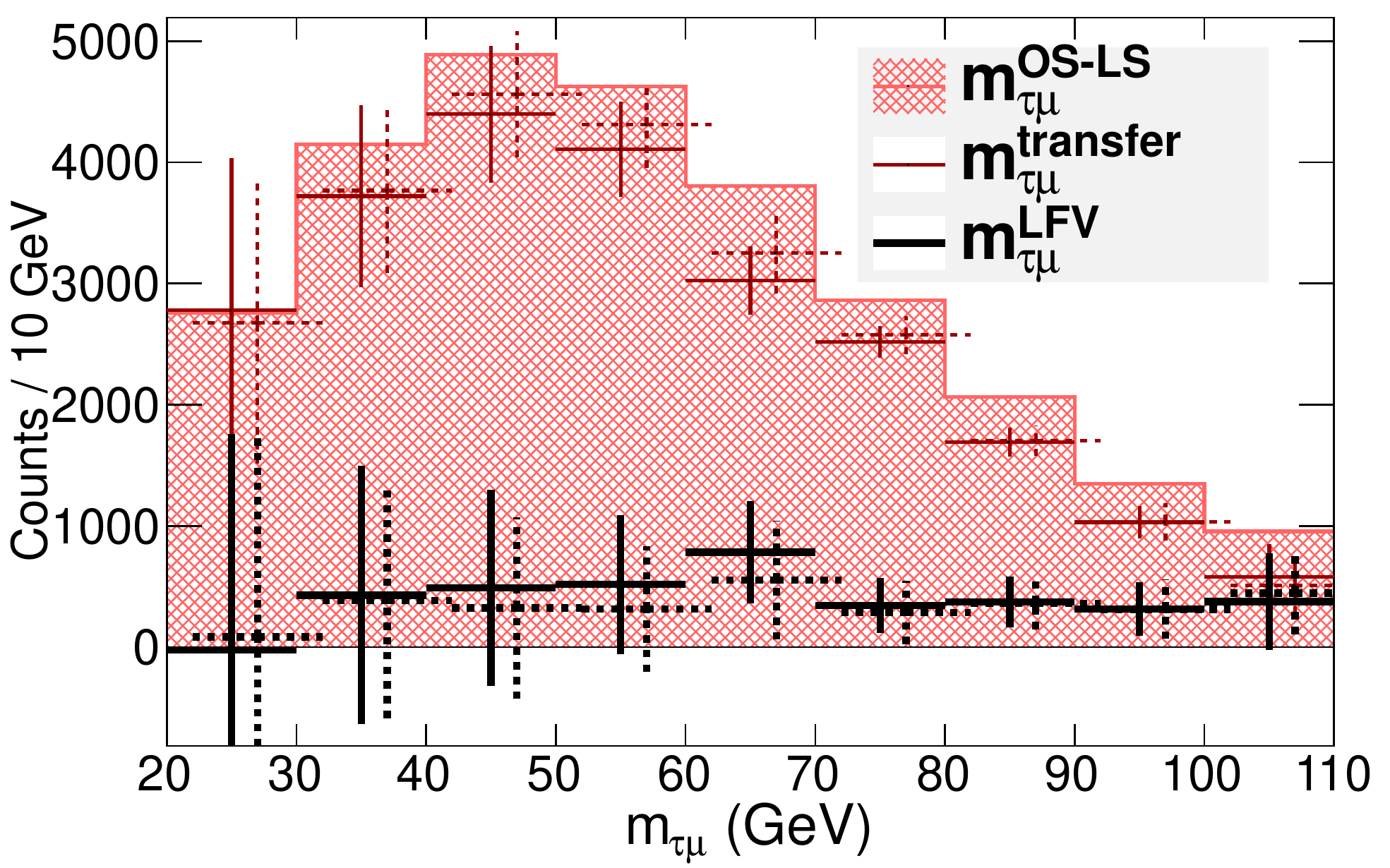}
	\caption{The $\tauh\mu$ invariant mass distribution for our LHC data $\deltaLFV=0.15$ point. The shape of the non-LFV distribution is estimated by using the transfer function as described in the text for the mass range $20~\gev < \mtaumu < 110~\gev$. This plot shows the result for both of the mass solutions shown in Table~\ref{tab:massMeasurements}. The second solution is dashed and shifted horizontally slightly for visibility. Note that the uncertainties shown for the resulting (black) $\mtaumu^{\rm{LFV}}$ histograms include our estimate of the systematic uncertainties. This result represents an integrated luminosity of $1000~\invfb$. However, we also report the situation for a lower luminosity of $300~\invfb$ in this paper.}
	\label{fig:deltaLFVfifteenResult}
\end{figure}

\begin{figure}
	\centering
	\includegraphics[width = 8cm]{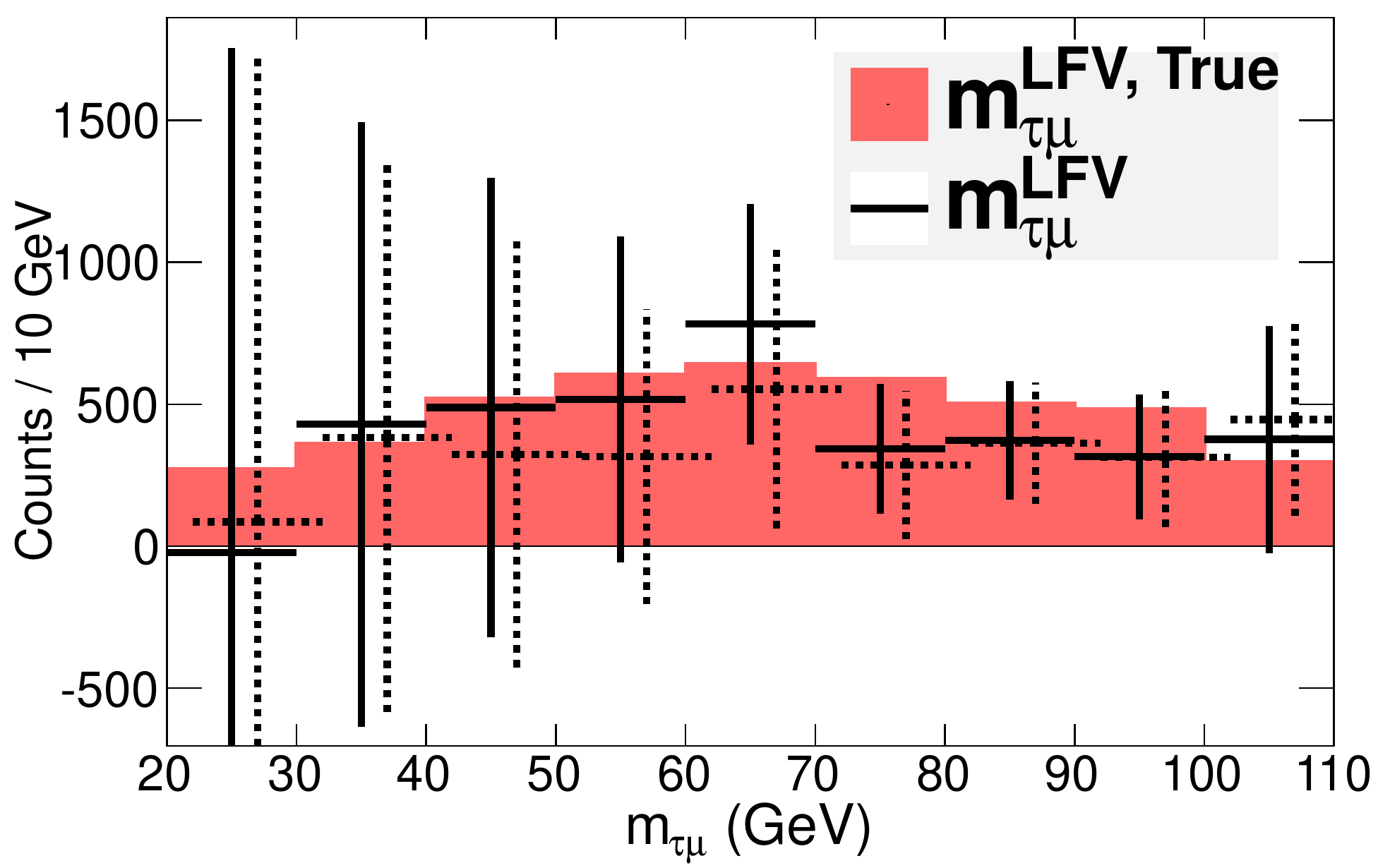}
	\caption{The black histogram shows the LFV $\tauh\mu$ invariant mass distribution for our LHC data $\deltaLFV=0.15$ point, compared to the filled red histogram, which is the Monte Carlo truth distribution of LFV decays for the same data set. This plot shows the result for both our mass solutions shown in Table~\ref{tab:massMeasurements}. The second solution is dashed and shifted horizontally slightly for visibility. Note that the uncertainties shown for the resulting (black) $\mtaumu^{\rm{LFV}}$ histograms include our estimate of the systematic uncertainties. This result represents an integrated luminosity of $1000~\invfb$. However, we also report the situation for a lower luminosity of $300~\invfb$ in this paper.}
	\label{fig:deltaLFVfifteenResultTruthCompare}
\end{figure}

We use this transfer function with our four nonzero LHC data $\deltaLFV$ points from Table~\ref{tab:deltaLFVChanges}. The result of this transfer function method is shown in Fig.~\ref{fig:deltaLFVfifteenResult}. This figure shows that this method has only a little sensitivity even for the largest allowed value of $\deltaLFV = 0.15$. This figure also shows that in spite of having two solutions for the measured SUSY masses, either mass solution will give us a similar result for the LFV signal. We compare the resulting $\mtaumu^{\rm{LFV}}$ shape of this analysis to the expected number of LFV events in Fig.~\ref{fig:deltaLFVfifteenResultTruthCompare}.

We determine how significant this signal is compared to the null case of $\deltaLFV = 0$ in the region in Fig.~\ref{fig:deltaLFVfifteenResult} where the uncertainties propagated from the transfer function are not too large, namely, $60~\gev < \mtaumu < 110~\gev$. We do this by assuming a Gaussian distribution for each bin based upon these uncertainties.
We find that the LFV signal for the first (second) SUSY mass solution shown in Table~\ref{tab:massMeasurements} has a $2.2\sigma$ ($1.6\sigma$) excess for a luminosity of $1000~\invfb$. We also note here the estimated luminosity requirements to obtain an equivalent signal for different values of $\deltaLFV$, which we show in Table~\ref{tab:luminosityReqs}. 
For a luminosity of $300~\invfb$,  this significance drops to $1.7\sigma$ ($1.2\sigma$). 
We note that the change in uncertainties on the SUSY mass and transfer function determinations do not scale as $\sqrt{\Lu}$ in our analysis. This indicates if the systematic uncertainty in our determination technique can be improved, the discovery potential is enhanced.

If there were no systematic uncertainty for measuring the $\tauh$ energy scale, the significance at $300~\invfb$ would bounce back to $2.2\sigma$ ($1.6\sigma$). Thus, a decrease in the systematic energy scale uncertainty at the LHC experiments would greatly improve the significance of this signal.

\begin{table}
\caption{The required luminosity to obtain an equivalent signal for different values of $\deltaLFV$ at our benchmark point. We estimate these luminosity values using our basic result of a $2.2\sigma$ ($1.6\sigma$) excess of LFV signal for our benchmark model with $\deltaLFV = 0.15$. We note that values of $\deltaLFV$ larger than $\sim15\%$ violate the bound on the branching ratio $\Br(\tau\rightarrow\mu\gamma)\leq 4.4 \times 10^{-8}$~\cite{BRtaumugamma}.}
\label{tab:luminosityReqs}
\begin{center}
\begin{tabular}{ccc} \hline \hline
    $\deltaLFV$(\%) & $\Br(\slepton{1} \rightarrow \mu \schizero{1})$ & $\Lu~(\invfb)$ \\ \hline
    5 & $3.1\times10^{-3}$ & $8390$ \\
    10 &  $1.2\times10^{-2}$ & $2170$ \\
    15 & $2.6\times10^{-2}$ & $1000$ \\
    32 & $1\times10^{-1}$ & $260$ \\
    45 & $2\times10^{-1}$ & $130$ \\
  \hline \hline
\end{tabular}
\end{center}
\end{table}

\section{Conclusions}\label{conclu}

In this paper, we investigated the possibility of finding evidence of LFV using a new technique, called a ``transfer function,'' along with mass reconstruction techniques at the LHC. 
We constructed the $\tauh\tauh$ and $\tauh\mu$ invariant mass distributions. We used the transfer function to convert the $\tauh\tauh$ mass distribution into the non-LFV $\tauh\mu$ mass distribution, $\mtaumu^{\rm{non-LFV}}$. The subtraction of the non-LFV $\tauh\mu$ distribution from the regular $\tauh\mu$ distribution left us with the LFV signal distribution, $\mtaumu^{\rm{LFV}}$. 

For our benchmark model ($\deltaLFV = 0.15$), shown in Table~\ref{tab:massBenchmark}, we needed $\sim300~\invfb$ to observe a $\sim2\sigma$ LFV signal for the case where $\Br(\slepton{1} \rightarrow \mu \schizero{1}) \simeq 3\%$. One can probe (at the $\sim2\sigma$ level) $\Br(\slepton{1} \rightarrow \mu \schizero{1})\simeq 20\%$ with a luminosity of 
$\sim45~\invfb$ for the parameter space discussed so far. However, these luminosity requirements are highly model dependent. If one goes to any other model point where $\schizero{2}$ decays into $\slepton{1} \tau \rightarrow \tau \tau \schizero{1}$, our analysis still applies. The luminosity requirement can be scaled by $\sigma_{\squark{},\gluino} \Br(\squark{},\gluino \rightarrow \schizero{2})$. One can reduce the requirement of luminosity with a reduction in the systematic uncertainty of the tau energy scale at the LHC experiments.

Lastly, we emphasize here that we have developed a new technique that can probe a LFV effect in this complex final state at the LHC. This technique will be equally effective to probe LFV in other SUSY models with a similar decay chain as a final state.

\section{Acknowledgements}\label{thanks}

We would like to give thanks to Sascha Bornhauser for his early work on this project, to Nathan Krislock for introducing us to the Nelder Mead method, to Kechen Wang and Kuver Sinha for their help in getting it to converge, and to Youngdo Oh for his helpful comments. This work is supported in part by DOE Grant No. DE-FG02-95ER40917 and by the World Class University (WCU) project through the National Research Foundation (NRF) of Korea funded by the Ministry of Education, Science, and Technology (Grant No. R32-2008-000-20001-0).


\begin{thebibliography}{20}
\bibitem{systematic}
The systematic uncertainties at the LHC will be evaluated correctly by the ATLAS and CMS collaborations. In this study, we demonstrate the effects of systematic uncertainties with a jet energy scale mismeasurement of 3\%. This mismeasurement is applied to jets, taus, and the missing transverse energy.
\end{thebibliography}
\end{document}